\documentclass[aps, prb, reprint, superscriptaddress, floatfix]{revtex4-2}
\usepackage{graphicx}
\usepackage{dcolumn}
\usepackage{bm}
\usepackage{physics}
\usepackage{amsfonts}
\usepackage{mathrsfs}
\usepackage{subfigure}
\usepackage[svgnames]{xcolor}
\usepackage[colorlinks=true,linkcolor=NavyBlue,anchorcolor=red,citecolor=NavyBlue, urlcolor=NavyBlue]{hyperref}
\usepackage{color}
\usepackage{xcolor}
\usepackage{ulem}

\DeclareMathAlphabet{\pazocal}{OMS}{zplm}{m}{n}
\usepackage{amsmath}
\usepackage{amssymb}
\usepackage[T1]{fontenc}
\begin{document}



\title{Emergent Spatial Textures from Interaction Quenches in the Hubbard Model}

\author{Sankha Subhra Bakshi}
\affiliation{Department of Physics, University of Virginia, Charlottesville, Virginia, 22904, USA}

\author{Gia-Wei Chern}
\affiliation{Department of Physics, University of Virginia, Charlottesville, Virginia, 22904, USA}

\date{\today}

\begin{abstract}
Interaction quenches in strongly correlated electron systems provide a powerful route to probe nonequilibrium many-body dynamics.
For the Hubbard model, nonequilibrium dynamical mean-field theory has revealed coherent post-quench oscillations, dynamical crossovers, and long-lived transient regimes. However, these studies are largely restricted to spatially homogeneous dynamics and therefore neglect the role of spatial structure formation during ultrafast evolution.
Here we investigate interaction quenches in the half-filled Hubbard model using a real-space time-dependent Gutzwiller framework.
We show that homogeneous nonequilibrium dynamics is generically unstable: even arbitrarily weak spatial fluctuations grow dynamically and drive the system toward intrinsically inhomogeneous states.
Depending on the interaction strength, the post-quench evolution exhibits spatial differentiation, nucleation, and slow coarsening of Mott-like domains.
Our results establish spatial self-organization as a generic feature of far-from-equilibrium correlated matter and reveal a fundamental limitation of spatially homogeneous nonequilibrium theories.
\end{abstract}

\maketitle

The nonequilibrium dynamics of strongly correlated electron systems have attracted intense interest over the past decade, driven largely by rapid advances in ultrafast experimental techniques~\cite{Rev1, Rev2, Rev3, Rev4}. Modern pump--probe and time-resolved spectroscopies now access femtosecond time scales, enabling direct observation of electronic, magnetic, and lattice dynamics far from equilibrium~\cite{trarpes,trrmxs,trtdts,trxrd2}. Importantly, nonequilibrium perturbations are no longer employed merely as diagnostic probes~\cite{LRT1}, but have emerged as powerful control knobs to engineer and manipulate quantum states of matter. 

While driven systems display rich phenomenology, a deeper theoretical understanding is facilitated by focusing on minimal nonequilibrium protocols that isolate intrinsic many-body dynamics. Quantum quenches, defined as sudden parameter changes, provide such a controlled framework~\cite{QQ3,QQ4,QQ5,QQ6}. By eliminating explicit driving fields, quenches directly expose fundamental nonequilibrium processes, including coherent many-body dynamics, relaxation pathways, dynamical criticality, and the emergence of nonthermal states. Among these, interaction quenches in the single-band Hubbard model~\cite{Hubbard} constitute a paradigmatic setting, as they probe the competition between kinetic energy and strong local correlations in its most direct form. These quenches are experimentally realizable in ultracold atomic systems via Feshbach tuning~\cite{UCAtoms1,UCAtoms2,UCAtoms3,UCAtoms4}, establishing them as both conceptually clean and experimentally relevant probes of nonequilibrium correlated matter.

From the theoretical perspective, one-dimensional systems are relatively well understood~\cite{ED1,ED2,tDMRG1,tDMRG2}. In higher dimensions, however, controlled treatments of nonequilibrium dynamics remain challenging. Time-dependent dynamical mean-field theory (DMFT) has emerged as a cornerstone approach~\cite{DMFT0,DMFT1,DMFT2,DMFT3,DMFT4,DMFT5,DMFT6}, capturing local quantum fluctuations exactly and revealing several robust features of interaction quenches, including nonequilibrium dynamical phase transitions, critical slowing down, and long-lived nonthermal states~\cite{Therm}.

Despite these successes, DMFT and its cluster extensions~\cite{ClusterDMFT1,DMFT5} fundamentally either neglect or strongly constrain spatial fluctuations. As a result, they are intrinsically incapable of describing the spontaneous emergence of spatial inhomogeneities, such as domains, defects, or textured order, following a quench. This omission is not merely technical. General arguments~\cite{KZ1,KZ2,KZ3} and experiments~\cite{KZexp1,KZexp2,KZexp3} indicate that quenches across or near instabilities naturally generate spatial structure. Moreover, a growing body of theoretical and numerical work demonstrates that quenched correlated systems---including superconducting, charge-ordered, and magnetically ordered phases---exhibit pronounced spatially inhomogeneous dynamics~\cite{chern19,huang19,yang25,Fan24,Lingyu2024,sankha2024,sankhaPRL2024}.

In the context of quench dynamics of Hubbard-type models, spatial inhomogeneities play a role in nonequilibrium dynamics that is conceptually analogous to that of quantum fluctuations: they provide essential relaxation channels and enable dephasing. At the same time, their inclusion allows for the emergence of nontrivial spatial textures that can qualitatively reshape post-quench evolution. Neglecting such effects can therefore lead to an incomplete, or even misleading, understanding of nonequilibrium behavior.

In this work, we address this gap by employing a real-space formulation of the time-dependent Gutzwiller approximation (TDGA) \cite{TDGA1,TDGA2,TDGA3}. While TDGA neglects dynamical quantum fluctuations, it offers a substantial computational advantage that enables large-scale simulations with spatially resolved variational parameters. This makes it particularly well suited for studying nonequilibrium dynamics in the presence of spatial degrees of freedom, including the spontaneous development of inhomogeneities, spatial dephasing, and the emergence of long-lived textures following interaction quenches \cite{TDGA4}.

Within this framework, allowing spatial degrees of freedom qualitatively changes the long-time quench dynamics. While the early-time response remains coherent and consistent with homogeneous TDGA and DMFT benchmarks, the subsequent evolution is governed by spatial dephasing between locally oscillating regions, providing an intrinsic relaxation channel even in the absence of dissipation \cite{TDGA1}. In the intermediate-coupling regime, this leads to the nucleation and slow coarsening of Mott-like regions, producing long-lived inhomogeneous states that are inaccessible within spatially uniform approaches.

We consider the single-band Hubbard model on a triangular lattice,
\begin{equation}
H(t)=\sum_{\langle ij\rangle,\sigma} -t_{ij}\, c_{i\sigma}^\dagger c_{j\sigma}
+ U(t)\sum_i n_{i\uparrow}n_{i\downarrow},
\end{equation}
where $c_{i\sigma}^\dagger$ ($c_{i\sigma}$) creates (annihilates) an electron with spin $\sigma$ at site $i$, $t_{ij}$ denotes nearest-neighbor hopping ($t_{\rm nn}$) and $U(t)$
is a time-dependent onsite interaction.
Within the time-dependent Gutzwiller approximation (TDGA)~\cite{DF1,SB1,SB2},
the many-body wavefunction is written as
$|\Psi_G(t)\rangle=\hat P_G(t)|\Psi_S(t)\rangle$, where $|\Psi_S(t)\rangle$
is a Slater determinant and $\hat P_G(t)$ is a product of local Gutzwiller
projectors parametrized by time-dependent variational amplitudes
$\Phi_i(t)$, which play the role of slave-boson fields. The resulting effective description naturally separates into two coupled
subsystems: quasiparticles evolving in the instantaneous background set by
$\Phi_i(t)$, which renormalizes the hopping and on-site potentials of the
effective quasiparticle Hamiltonian $H_{\mathrm{qp}}$, and local Gutzwiller
amplitudes $\Phi_i(t)$ that evolve under the feedback of the quasiparticle
state through an effective slave-boson Hamiltonian $H_{\mathrm{sb}}$. At any time, the quasiparticle density matrix
$\rho_{j\sigma,i\sigma'}=\langle c^\dagger_{i\sigma'}c_{j\sigma}\rangle$
obeys a von~Neumann equation,
\begin{equation}
\frac{d\rho}{dt}
=
i[\rho,H_{\mathrm{qp}}(\{\Phi_i\})],
\label{eq:rho_main}
\end{equation}
while the local Gutzwiller amplitudes evolve according to
\begin{equation}
i\frac{d\Phi_i}{dt}
=
\frac{\partial}{\partial \Phi_i^\dagger}
H_{\mathrm{sb}}\left(\{\Phi_i\}, \rho\right).
\label{eq:phi_main}
\end{equation}
These equations are solved simultaneously in real time, ensuring full
self-consistency between quasiparticle motion and local correlation dynamics.
We refer to this real-space implementation as the Gutzwiller von~Neumann
dynamics (GvND) method.
Explicit expressions and numerical details are provided in the Appendix~\ref{app:gvnd}.

\begin{figure}[t!]
\includegraphics[width=5cm, height=7cm]{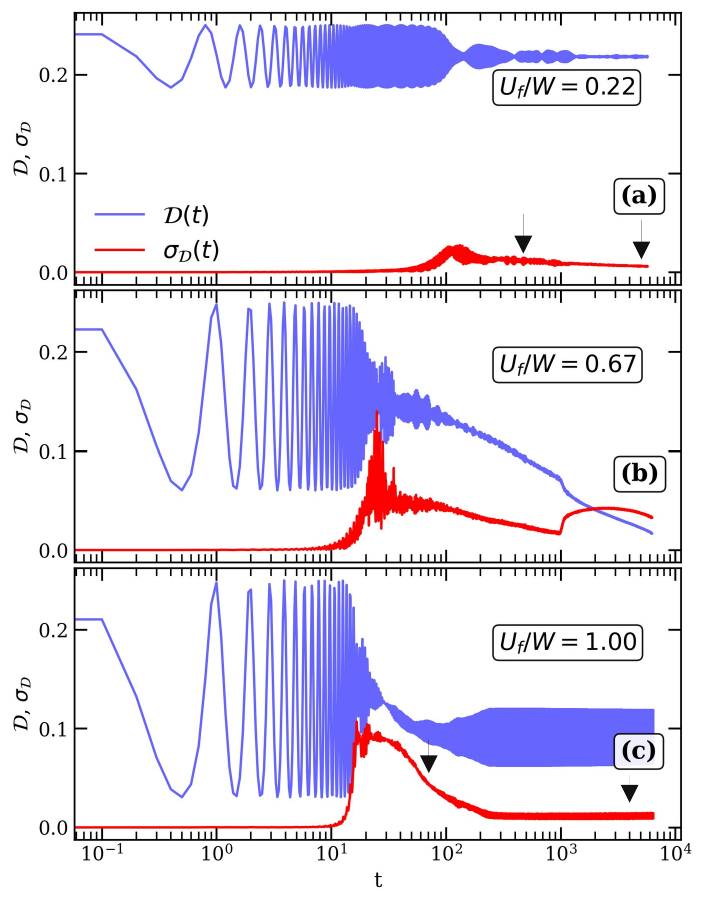}
\includegraphics[width=3.5cm, height=7cm]{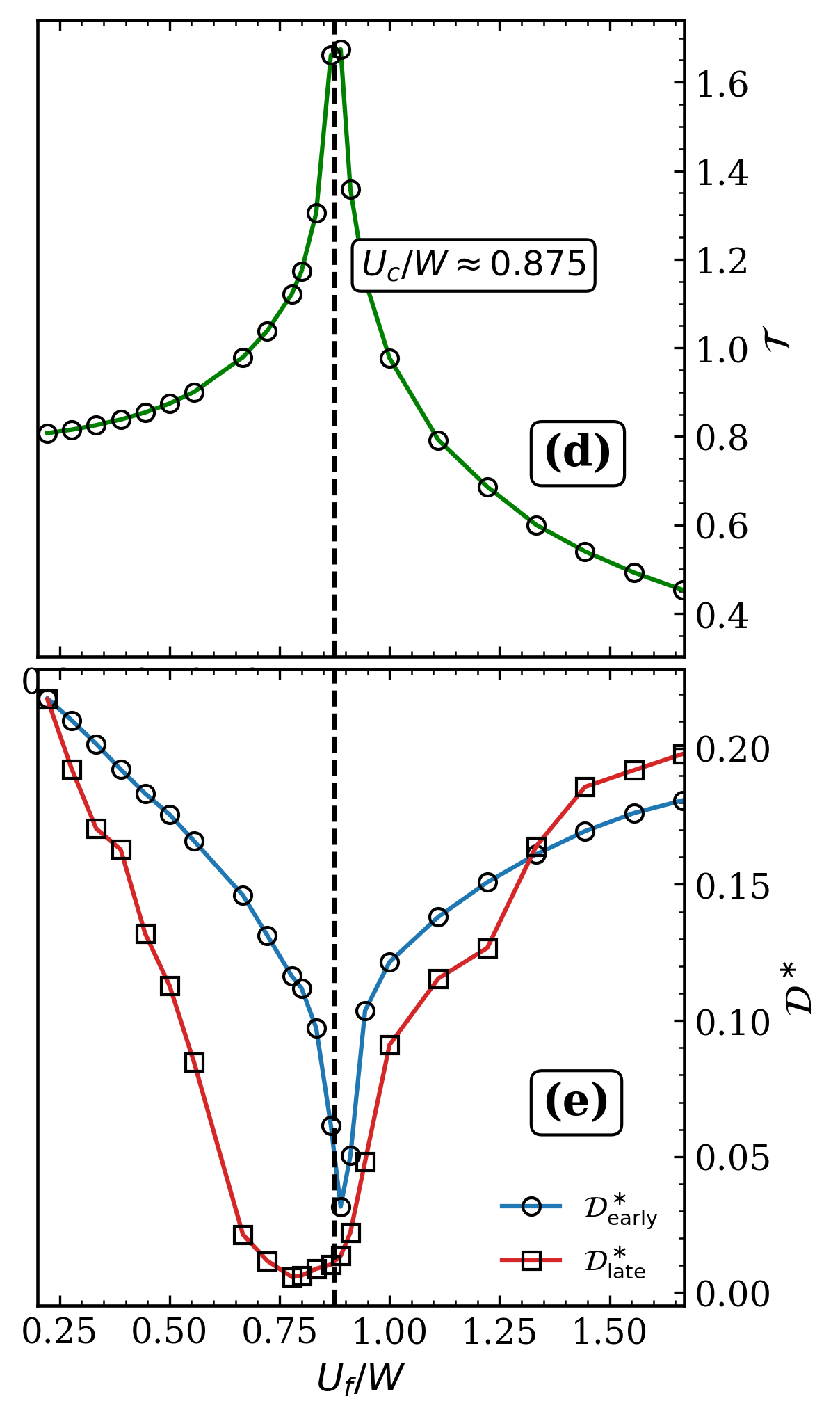}
\caption{%
GvND dynamics following an interaction quench to final strength $U_f$.
Panels (a)--(c) show the time evolution of the double occupancy
$\mathcal{D}(t)$ and its spatial standard deviation $\sigma_{\mathcal{D}}(t)$
for weak, intermediate, and strong quenches.
Panel (d) shows the oscillation period $\mathcal{T}$ extracted from the
early-time dynamics as a function of $U_f/W$.
Panel (e) shows the early-time and long-time averaged double occupancy
$D^*$ with varying $U_f$.
}

\label{fig:gvnd_quench}
\end{figure}

We restrict ourselves to the paramagnetic regime. In this setting, the local Gutzwiller amplitudes can be taken in diagonal form,
$
\Phi_i(t)=\mathrm{diag}(e_i,p_i,d_i),
$
corresponding to empty, singly occupied, and doubly occupied local configurations, respectively. The system is initialized in the noninteracting ground state at zero temperature ($T=0$) with interaction strength $U_i=0$. At time $t=0$, the interaction is suddenly quenched to a finite value $U_f$, and the system subsequently undergoes unitary time evolution governed by the interacting Hamiltonian.

To allow for the emergence of spatial dynamics without explicitly breaking symmetries, we introduce a weak on-site Anderson disorder potential ${\epsilon_i}$ with magnitude $10^{-5}t_{nn}$. All simulations are performed within the paramagnetic GvND framework on a $48\times48$ lattice. Time is measured in units of $\hbar/t_{nn}$ throughout. We monitor the local density $n_i=|p_i|^2+|d_i|^2$ and the local double occupancy $\mathcal{D}_i(t)=|d_i(t)|^2$. Details regarding timestep stability and numerical convergence are provided in the Appendix~\ref{app:dt}.

We characterize the quench dynamics through the spatially averaged double occupancy $\mathcal{D}(t)$, shown in Fig.~\ref{fig:gvnd_quench}(a–c) for representative weak-, intermediate-, and strong-coupling interaction quenches. For all final interaction strengths, the evolution of $\mathcal{D}(t)$ exhibits a clear two-stage structure: an initial regime of coherent collective dynamics, followed at longer times by a qualitatively distinct regime dominated by spatial dephasing.

In the early-time regime, $\mathcal{D}(t)$ displays coherent oscillations characterized by a quench-dependent period $\mathcal{T}$ and a reduced mean value $\mathcal{D}^\ast$. This early-time behavior coincides with that obtained within homogeneous TDGA~\cite{TDGA1}, reflecting a collective quasiparticle mode of the correlated state. The oscillation period displays a pronounced maximum as a function of the final interaction strength $U_f$ [Fig.~\ref{fig:gvnd_quench}(d)], while the early-time averaged value $\mathcal{D}^\ast$ varies nonmonotonically with $U_f$ [Fig.~\ref{fig:gvnd_quench}(e)], consistent with proximity to a dynamical critical point $U_c$. This provides our reference scale for classifying quenches as weak ($U \ll U_c$), intermediate ($U \approx U_c$), and strong ($U \gg U_c$). The redistribution of spectral weight and the associated renormalization of the effective bandwidth are discussed in the Appendix~\ref{app:SW}.

We note that the inclusion of local quantum fluctuations can lead to damping of these early-time oscillations, as demonstrated in nonequilibrium DMFT simulations~\cite{Therm}.  Such fluctuation-induced  decoherence is expected to act as a seed for the subsequent emergence of spatial inhomogeneities, which we analyze below with the GvND framework.

\begin{figure}[t]
    \raggedleft
    \includegraphics[width=8.5cm, height=4.5 cm]{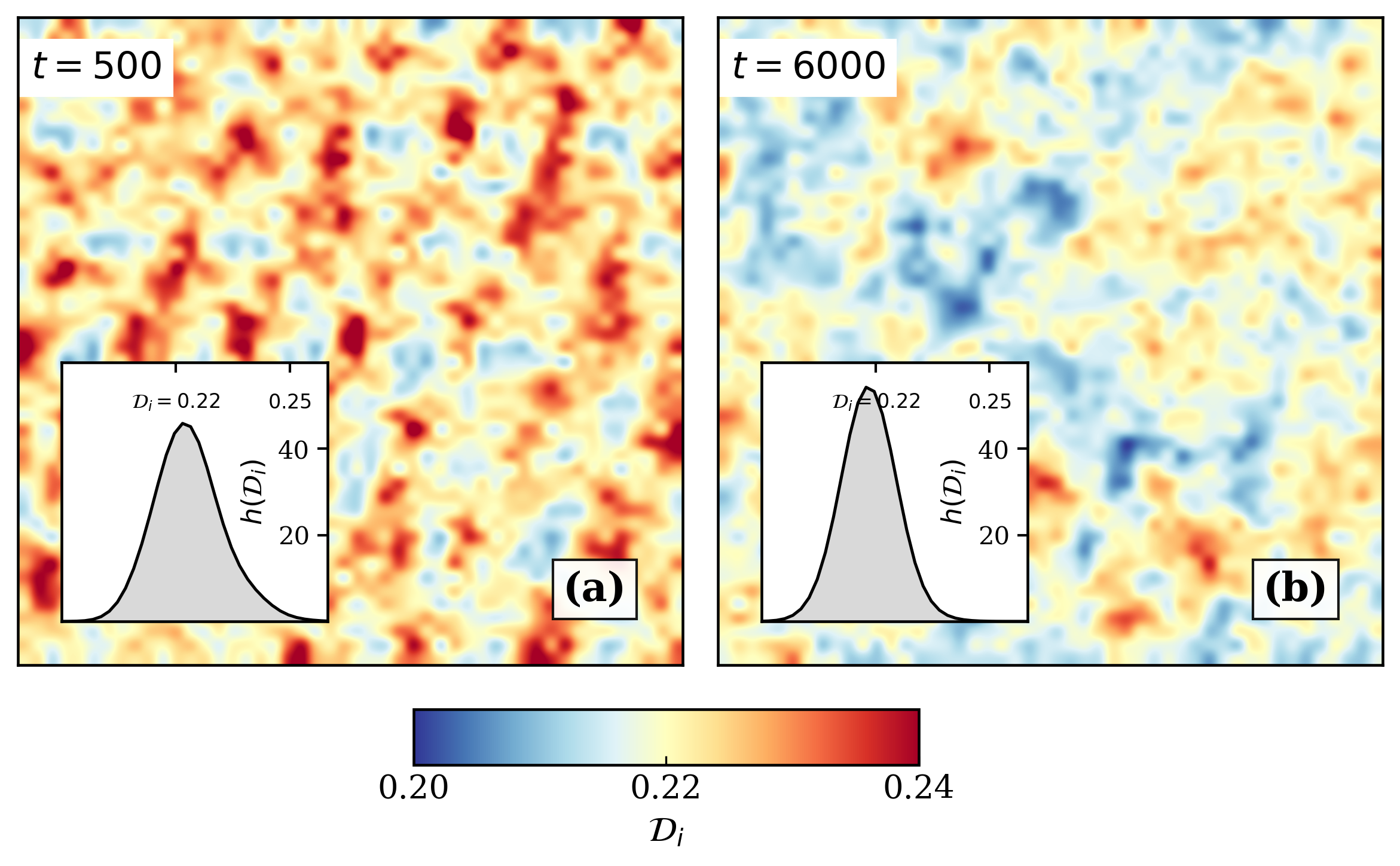}
    \caption{%
    Snapshots of the spatial distribution of the double occupancy
    $\mathcal{D}(x,y)$ at intermediate time $t=500$ and long time $t=6000$ for interaction strength $U_f/W=0.22$. The corresponding distributions of $\mathcal{D}$ are shown in the insets.
    }
    \label{fig:fig2}
\end{figure}

At longer times, the post-quench dynamics enters a regime that departs qualitatively from the homogeneous description. The global oscillations of $\mathcal{D}(t)$ collapse beyond a characteristic timescale, accompanied by a rapid growth of the spatial standard deviation $\sigma_{\mathcal{D}}(t)$. This crossover marks the onset of spatial phase decoherence and the spontaneous development of inhomogeneity. Unlike in DMFT, the damping observed here originates from dephasing between locally oscillating regions rather than from local quantum dissipation. A detailed discussion of the interaction dependence of the early- and long-time regimes is provided in the Appendix~\ref{app:regimes}.

To elucidate the nature of the emergent inhomogeneity, we examine snapshots of
the local double occupancy $\mathcal{D}(x,y)$ at the times indicated by arrows
in Fig.~\ref{fig:gvnd_quench} for weak and strong quenches. Fig.~\ref{fig:fig2} shows
snapshots at intermediate and long times following a representative weak
quench. At intermediate times, the system exhibits weak spatial modulations with
approximately repeating patterns and small amplitude
[Fig.~\ref{fig:fig2}(a)]. The corresponding distribution of $\mathcal{D}$
roughly assumes a Gaussian form. At long times, these modulations are washed out
and no clear spatial correlations remain. The system evolves toward an
approximately homogeneous quasi-stationary state, and the distribution of
$\mathcal{D}$ becomes a slightly narrower Gaussian
[Fig.~\ref{fig:fig2}(b)], consistent with spatial dephasing and the randomization
of local phases.

\begin{figure}[t]
    \includegraphics[width=8.5cm, height=4.5 cm]{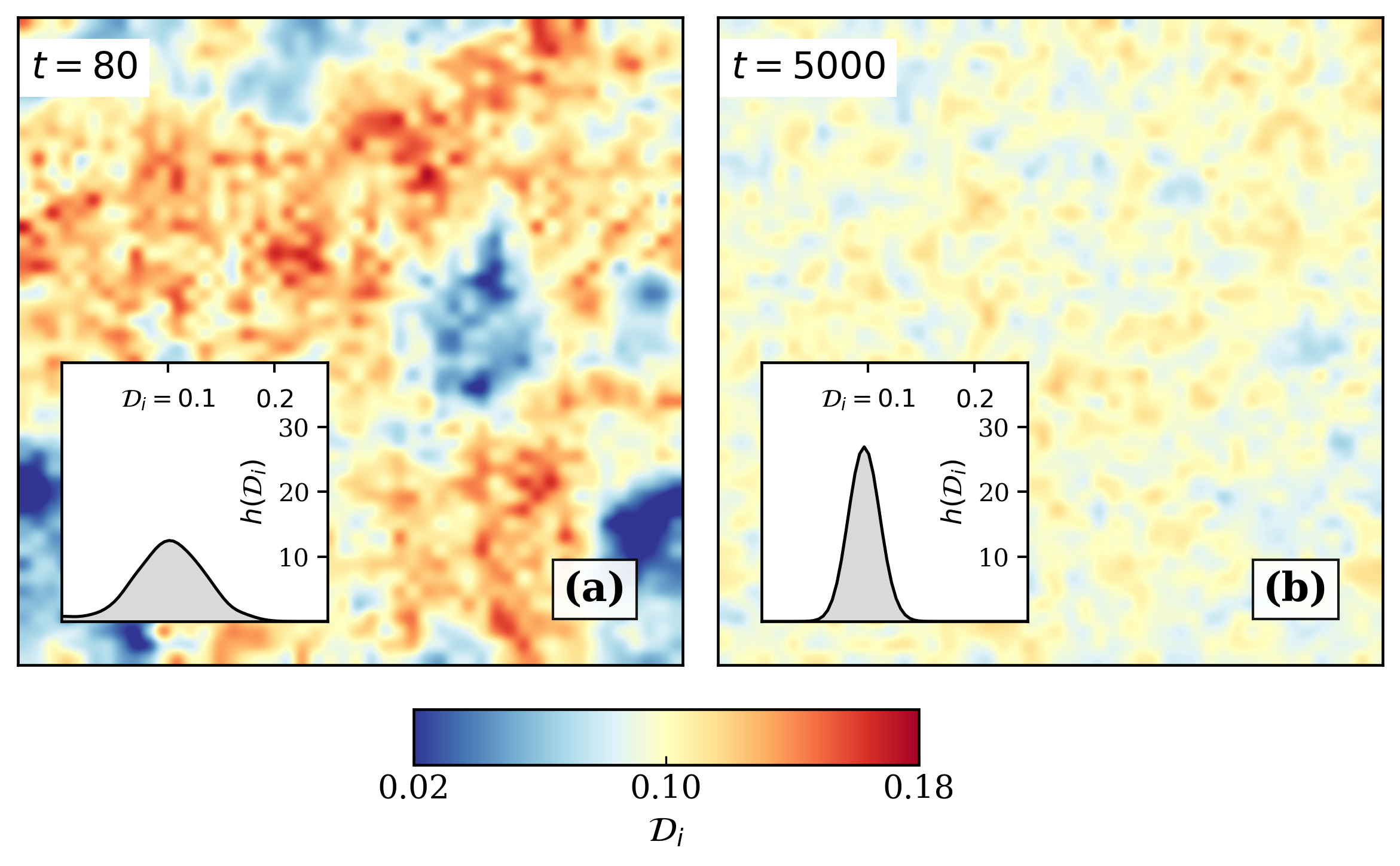}
\caption{%
Snapshots of the spatial distribution of the double occupancy
$\mathcal{D}(x,y)$ at intermediate time $t=80$ and long time $t=5000$ for interaction strength $U_f/W=1.0$. The corresponding distributions of $\mathcal{D}$ are shown in the insets.
}
    \label{fig:fig3}
\end{figure}

For stronger quenches, the spatial dynamics changes qualitatively. The
spatially averaged double occupancy $\mathcal{D}(t)$ initially exhibits
coherent oscillations, followed by a loss of global coherence as the spatial
fluctuation $\sigma_{\mathcal{D}}(t)$ grows, and eventually a partial recovery
of coherence at long times as $\sigma_{\mathcal{D}}(t)$ decreases. We focus on
two representative stages of this evolution: an intermediate time after the
loss of coherence and the long-time regime. As shown in Fig.~\ref{fig:fig3}(a), at intermediate times the system develops
extended regions with strongly suppressed double occupancy coexisting with
more itinerant regions, resulting in broad and distinctly non-Gaussian
distributions of $\mathcal{D}_i$. These inhomogeneities arise dynamically from
unitary time evolution and are not induced by any externally imposed
symmetry-breaking field. At long times, the spatial inhomogeneities are substantially reduced and the
double occupancy recovers a finite value with a narrow, approximately Gaussian
distribution [Fig.~\ref{fig:fig3}(b)]. Although this long-time state exhibits
qualitative similarities to the early-time coherent regime, its properties
are strongly renormalized by the transient inhomogeneous dynamics encountered
at intermediate times. 

\begin{figure}[t!]
    \includegraphics[width=8.7cm, height=2.7cm]{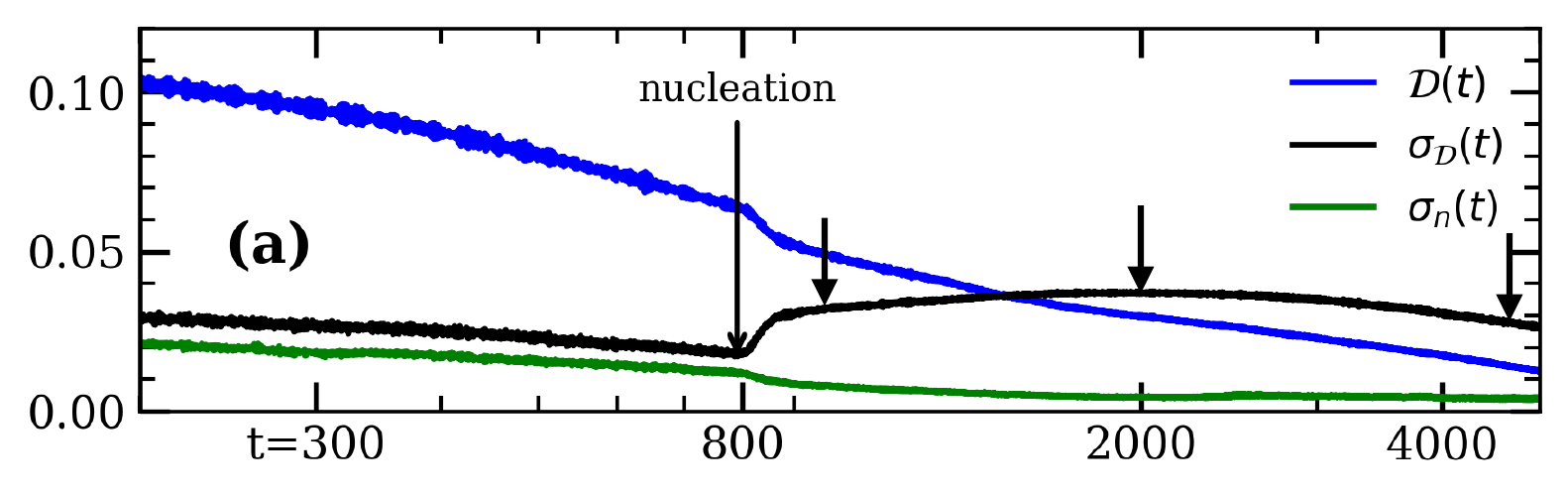}
    \raggedleft
    \includegraphics[width=8cm, height=7.5cm]{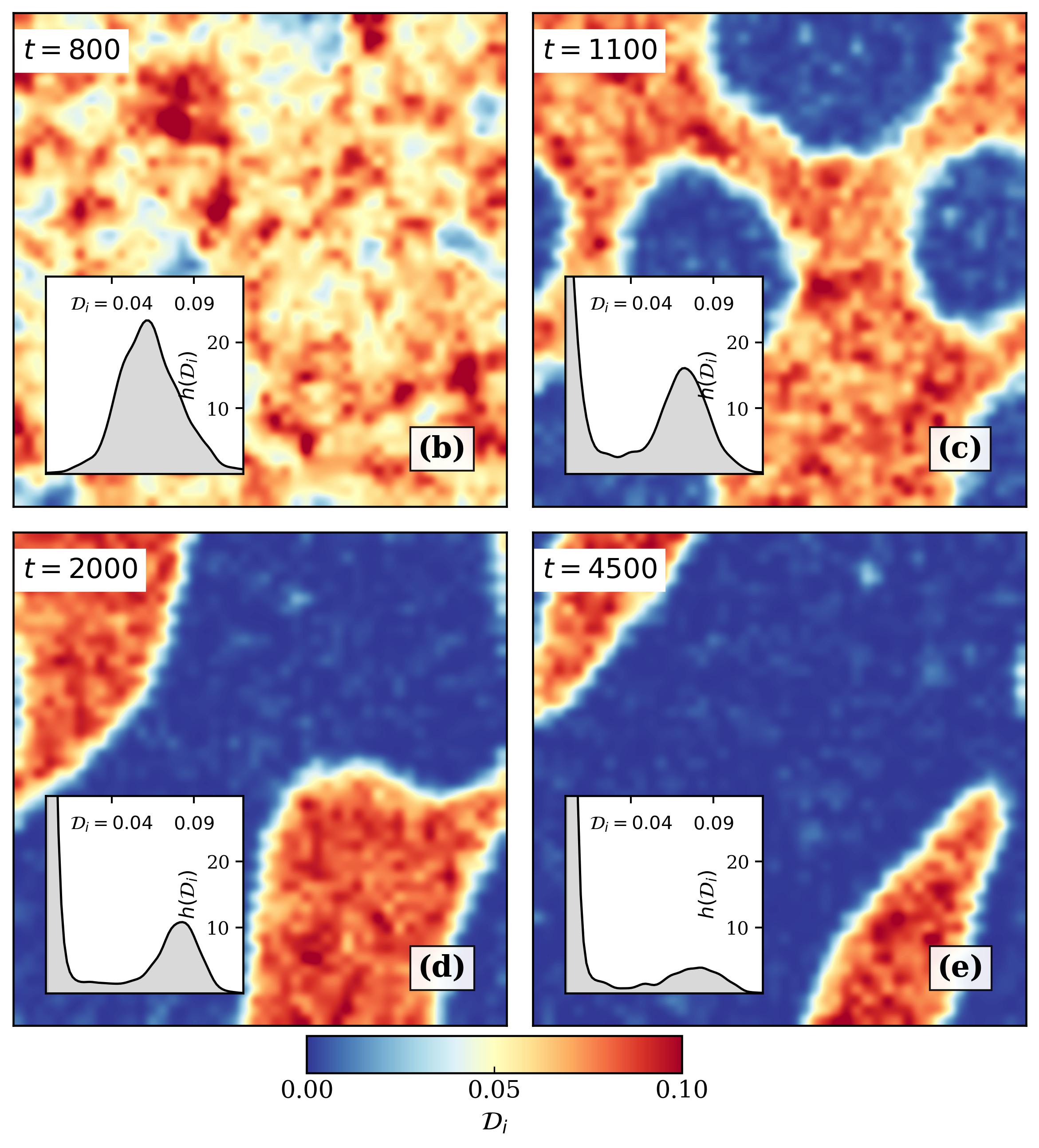}
\caption{%
Coarsening dynamics following a quench to $U_f/W=0.72$.
(a) Time evolution of the spatially averaged double occupancy
$\mathcal{D}(t)$, its spatial standard deviation $\sigma_{\mathcal{D}}(t)$,
and the density fluctuation $\sigma_n(t)$.
(b)–(e) Snapshots of the local double occupancy $\mathcal{D}_i$ at the
indicated times, showing domain nucleation followed by coarsening.
Insets show the corresponding distributions of $\mathcal{D}_i$.
}
    \label{fig:fig4}
\end{figure}

The most pronounced deviations from homogeneous dynamics occur in the
intermediate-coupling regime, where spatial inhomogeneities play a decisive
role. While the qualitative growth and decay of
$\sigma_{\mathcal{D}}(t)$ are common to all interaction strengths, only in the
intermediate regime do these fluctuations lead to nontrivial long-time
dynamics. As shown in Fig.~\ref{fig:fig4}(a), the decay of the spatially
averaged double occupancy around $t\simeq 800$ coincides with a sharp increase
in $\sigma_{\mathcal{D}}(t)$, signaling the onset of nucleation. In contrast,
the charge density fluctuation $\sigma_n(t)$ decreases monotonically, indicating that
the emerging structures are driven by correlation effects rather than charge
redistribution. Snapshots of the local double occupancy $\mathcal{D}_i$, shown in
Fig.~\ref{fig:fig4}(b--e) and taken at the times marked by arrows in
Fig.~\ref{fig:fig4}(a), reveal the formation of Mott-like regions that grow and
merge over long timescales. Correspondingly, the probability distributions
evolve from a broad single peak to a bimodal form dominated by a growing peak
near $\mathcal{D}\approx 0$, providing clear evidence for phase coexistence and
slow coarsening.

\begin{figure}[t!]
    \centering
    \includegraphics[width=\linewidth]{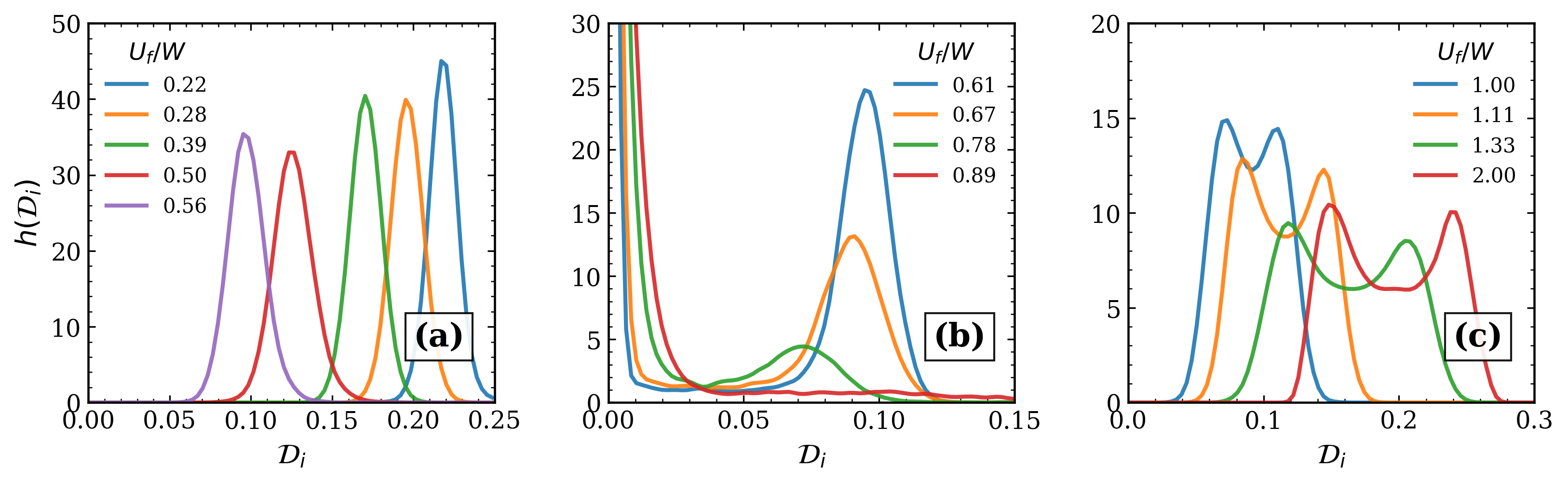}
\caption{%
Probability distributions of the local double occupancy $\mathcal{D}_i(t)$
constructed from all sites of a $48\times48$ lattice and averaged over the
time window $t\in[1000,1100]$. Panels (a)–(c) correspond to quenches in the weak-, intermediate-, and
strong-coupling regimes, respectively.
}
    \label{fig:fig5}
\end{figure}

Further support for this interpretation is provided by the probability
distributions of the local double occupancy $\mathcal{D}_i(t)$, shown in
Fig.~\ref{fig:fig5} for intermediate time windows. In the weak-coupling regime,
the distribution is broader than a Gaussian, reflecting finite spatial
fluctuations. In contrast, for strong quenches the instantaneous distributions
remain narrow and approximately Gaussian; however, temporal oscillations
of their mean lead to an effectively broadened, overlapping distribution upon
time averaging. By comparison, the intermediate-coupling regime exhibits a
distinctly bimodal structure, providing clear statistical evidence for the
coexistence of Mott-like and itinerant regions during the post-quench
evolution.

Taken together, Figs.~\ref{fig:fig4} and \ref{fig:fig5} demonstrate that near
the dynamical instability spatial inhomogeneities play a central role in
shaping the long-time dynamics. The evolution proceeds via nucleation and
coarsening rather than smooth relaxation, highlighting an intrinsic
nonequilibrium instability of homogeneous solutions once spatial degrees of
freedom are included, even in the absence of explicit dissipation.

The dynamical regimes identified here are separated by crossovers rather than
sharp phase transitions, characterized by interaction-dependent timescales
associated with the loss of global coherence and the growth of spatial
inhomogeneities. While full thermalization is not expected within the GA
framework, inclusion of spatial fluctuations opens additional relaxation
channels through spatial dephasing and redistribution of correlations. This
suggests a route toward effective equilibration that is absent in spatially
uniform treatments. The associated timescales may exhibit system-size
dependence, as commonly observed in coarsening dynamics; a systematic analysis
of finite-size effects is presented in the Appendix~\ref{app:size},
but a detailed scaling study is beyond the scope of the present work.

In summary, we have employed a real-space Gutzwiller–von Neumann dynamics
approach to study interaction quenches in the half-filled Hubbard model on a
triangular lattice. We find that the synchronized oscillatory dynamics predicted
by spatially homogeneous theories is intrinsically unstable: even weak
inhomogeneities are dynamically amplified, leading to spatial dephasing and the
emergence of textured nonequilibrium states. Depending on the interaction
strength, the post-quench evolution separates into three distinct dynamical
regimes: weak quenches lead to nearly homogeneous relaxation, intermediate
quenches exhibit nucleation and slow coarsening of Mott-like regions, and
strong quenches display a collapse followed by a partial re-emergence of global
coherence.

Our results highlight a complementary perspective to nonequilibrium DMFT.
While DMFT emphasizes the role of local quantum fluctuations encoded along the
imaginary-time axis, we demonstrate that purely real-space fluctuations can
produce qualitatively similar relaxation phenomena through spatial dephasing
and self-organization. An important open direction is to develop approaches
that incorporate both spatial and quantum fluctuations on equal footing. More
generally, our findings show that spatial self-organization is a key ingredient
of far-from-equilibrium dynamics in correlated systems and must be explicitly
accounted for in theoretical descriptions of driven quantum matter.

\begin{acknowledgments}
This work was supported by the Owens Family Foundation. The authors acknowledge Research Computing at the University of Virginia
for providing computational resources and technical support.
\end{acknowledgments}

\clearpage

\appendix

\section{Real-Space Gutzwiller--von Neumann Dynamics Formalism}
\label{app:gvnd}

In this section we provide a detailed description of the real-space
Gutzwiller--von Neumann dynamics (GvND) formalism employed in the main text. We consider the single-band Hubbard model
\begin{equation}
\hat H(t)
=
\sum_{ij,\sigma}
-t_{ij} \hat c^{\dagger}_{i\sigma} \hat c_{j\sigma}
+
U(t) \sum_i \hat n_{i\uparrow} \hat n_{i\downarrow},
\label{eq:hubbard}
\end{equation}
where $t_{ij}$ denotes the hopping amplitude between sites $i$ and $j$,
$U(t)$ is the (possibly time-dependent) onsite interaction, and
$\hat n_{i\sigma}=\hat c^{\dagger}_{i\sigma}\hat c_{i\sigma}$.

The time-dependent Gutzwiller wave function is written as
\begin{equation}
\ket{\Psi_G(t)}=\hat P_G(t)\ket{\Psi_S(t)},
\end{equation}
where $\ket{\Psi_S(t)}$ is a Slater determinant describing itinerant
quasiparticles and $\hat P_G(t)=\prod_i \hat P_i(t)$ is a product of local
Gutzwiller projectors encoding correlation effects. Each local projector
$\hat P_i$ acts in the local Hilbert space
\(
\{\ket{0},\ket{\uparrow},\ket{\downarrow},\ket{\uparrow\downarrow}\}
\)
and is parametrized by a site-dependent variational matrix $\Phi_i$.

This formulation is formally equivalent to the rotationally invariant
slave-boson representation of correlated electron systems~\cite{SB1,TDGA4}. Within this representation, expectation values
of local operators reduce to traces over the local Hilbert space,
\begin{equation}
\langle \Psi_G | \hat O_i | \Psi_G \rangle
=
\mathrm{Tr}
\!\left(
\hat O_i \, \Phi_i^\dagger \Phi_i
\right).
\end{equation}

Throughout this work we restrict to diagonal local correlators,
\begin{equation}
\hat P_i
=
\sum_{\Gamma}
\Phi_{i\Gamma}\ket{\Gamma}_i\bra{\Gamma},
\end{equation}
where $\Gamma$ labels the four local charge configurations and the complex
amplitudes $\Phi_{i\Gamma}$ control their statistical weights.

The equations of motion follow from the Dirac--Frenkel time-dependent
variational principle~\cite{Dirac1930,DF1}, applied to the action
\begin{equation}
\mathcal{S}
=
\int dt\,
\mel{\Psi_G(t)}{i\partial_t-\hat H(t)}{\Psi_G(t)}.
\end{equation}
Independent variation with respect to $\ket{\Psi_S}$ and $\Phi_i$ yields a
closed set of coupled dynamical equations~\cite{TDGA1,TDGA4}:
\begin{align}
i\partial_t \ket{\Psi_S}
&=
\hat H_{\mathrm{GA}}[\{\Phi_i\}]\ket{\Psi_S},
\label{eq:slater_dyn}
\\
i\partial_t \Phi_i
&=
\frac{\partial}{\partial \Phi_i^\dagger}
\expval{\hat H_{\mathrm{GA}}},
\label{eq:phi_dyn}
\end{align}
where $\hat H_{\mathrm{GA}}$ is the effective Gutzwiller Hamiltonian.

Within the Gutzwiller approximation, the latter takes the form
\begin{equation}
\hat H_{\mathrm{GA}}
=
\sum_{ij,\sigma}
-t_{ij} R_{i\sigma}R^{\ast}_{j\sigma}
\hat c^{\dagger}_{i\sigma}\hat c_{j\sigma}
+
\sum_i U \hat D_i
+
\sum_{i\sigma}\mu_{i\sigma}\hat n_{i\sigma},
\label{eq:hga}
\end{equation}
where $\hat D_i=\mathrm{diag}(0,0,0,1)$ is the local double-occupancy operator
and $\mu_{i\sigma}$ are Lagrange multipliers enforcing local charge
constraints.

The hopping renormalization matrices $R_{i\sigma}$ encode the feedback of local
correlations on quasiparticle motion and are defined as
\begin{equation}
R_{i,\alpha\beta}
=
\frac{
\mathrm{Tr}
\!\left(
\Phi_i^\dagger
\hat c^{\dagger}_{i\alpha}
\Phi_i
\hat c_{i\beta}
\right)
}
{\sqrt{n_{i\beta}(1-n_{i\beta})}},
\label{eq:Rdef}
\end{equation}
which for the single-band Hubbard model reduces to
\begin{equation}
R_{i,\sigma}
=
\frac{
\Phi_{i0}^\ast \Phi_{i\sigma}
+
\Phi_{i\bar\sigma}^\ast \Phi_{i\uparrow\downarrow}
}
{\sqrt{n_{i\sigma}(1-n_{i\sigma})}}.
\label{eq:Rexplicit}
\end{equation}

We introduce the single-particle density matrix
\begin{equation}
\rho_{j\sigma,i\sigma'}
=
\mel{\Psi_S}{\hat c^{\dagger}_{i\sigma'}\hat c_{j\sigma}}{\Psi_S},
\end{equation}
whose time evolution obeys a von~Neumann equation,
\begin{equation}
\frac{d\rho}{dt}
=
i[\rho, H_{\mathrm{qp}}],
\end{equation}
where,
\begin{equation}
[H_{\text{qp}}]_{i\sigma, j\sigma'}
= - t_{i\sigma, j\sigma'}R_{i\sigma} R^\ast_{j\sigma'}+\mu_{i\sigma}\delta_{ij}\delta_{\sigma,\sigma'}
\end{equation}

\begin{figure}[t!]
    \centering
    \includegraphics[width=0.85\linewidth]{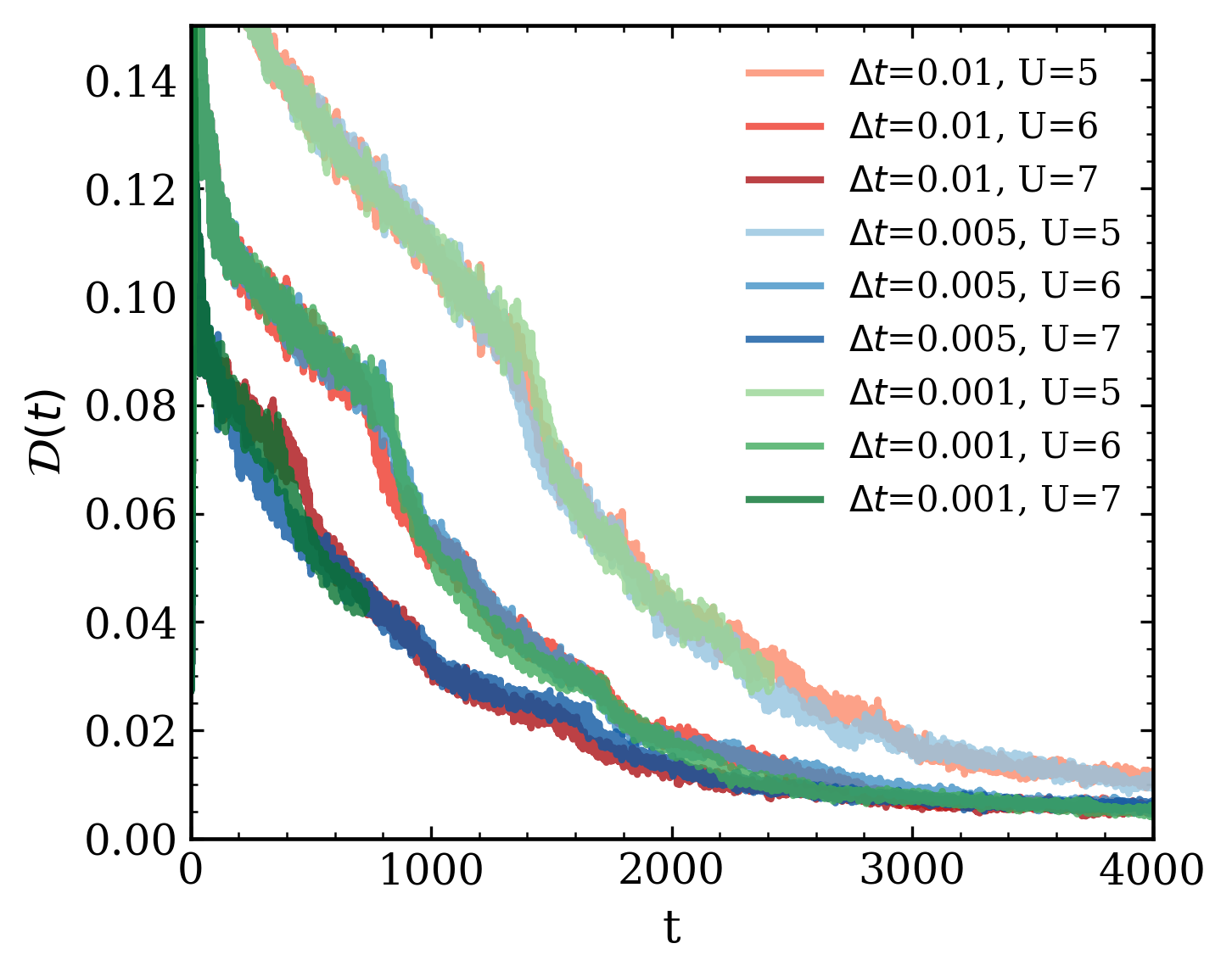}
\caption{%
Time evolution of the double occupancy $\mathcal{D}(t)$ obtained using different timesteps
$\Delta t=0.001$, $0.005$, and $0.01$ for several $U$ values, demonstrating numerical stability.
}
    \label{fig:dt}
\end{figure}
which constitutes the quasiparticle sector of the GvND dynamics
\cite{TDGA5}. Explicitly,
\begin{align}
\frac{d \rho_{i\sigma,j\sigma}}{dt}
&=
i(\mu_{j\sigma}-\mu_{i\sigma})\rho_{i\sigma,j\sigma}
\nonumber\\
&\quad-
i\sum_k
\left(
t_{ik}R_{i\sigma}R_{k\sigma}^\ast\rho_{k\sigma,j\sigma}
-
t_{kj}R_{k\sigma}R_{j\sigma}^\ast\rho_{i\sigma,k\sigma}
\right).
\label{eq:rho_dyn}
\end{align}

\begin{figure}[t!]
\includegraphics[width=5cm, height=7cm]{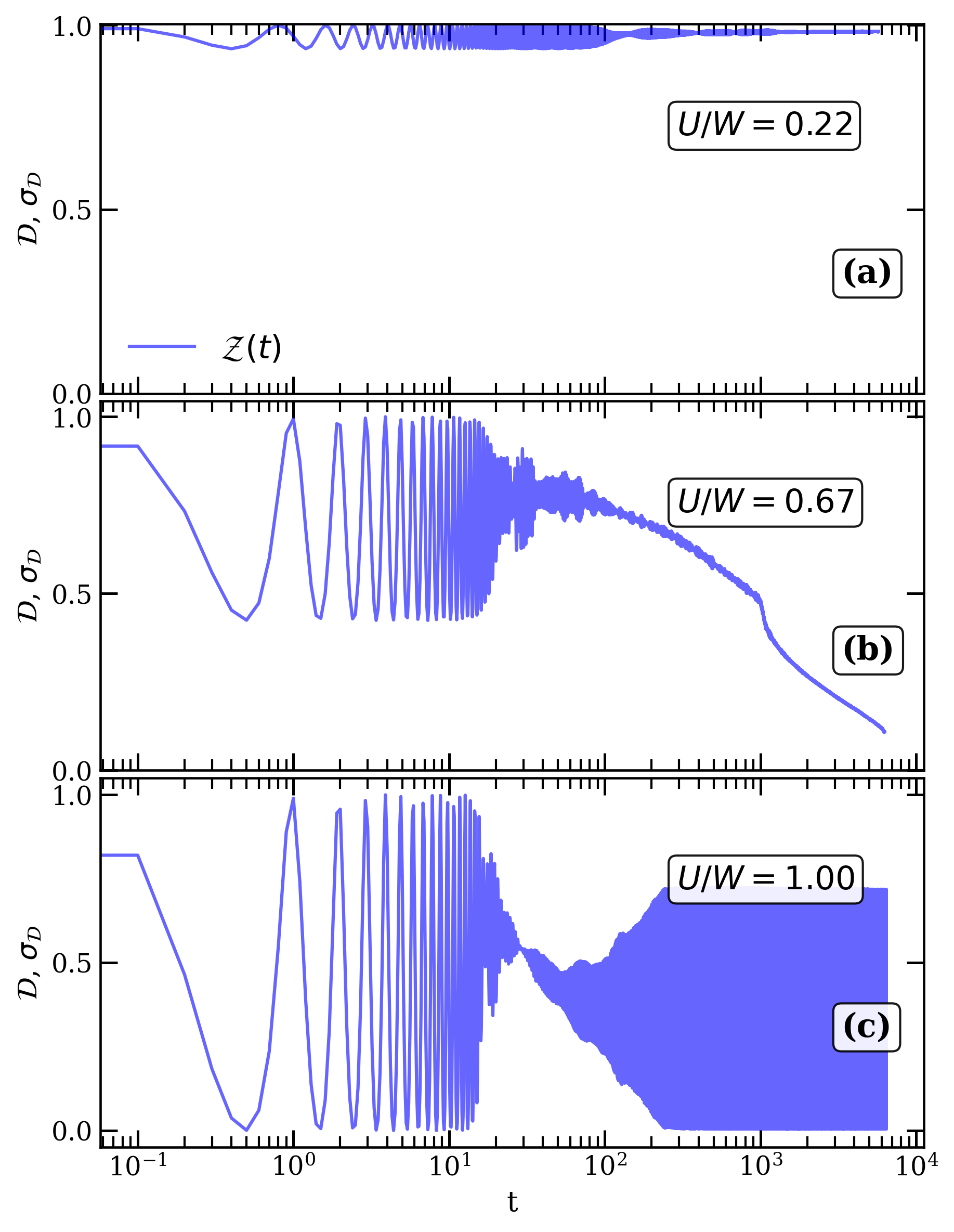}
\includegraphics[width=3.5cm, height=7cm]{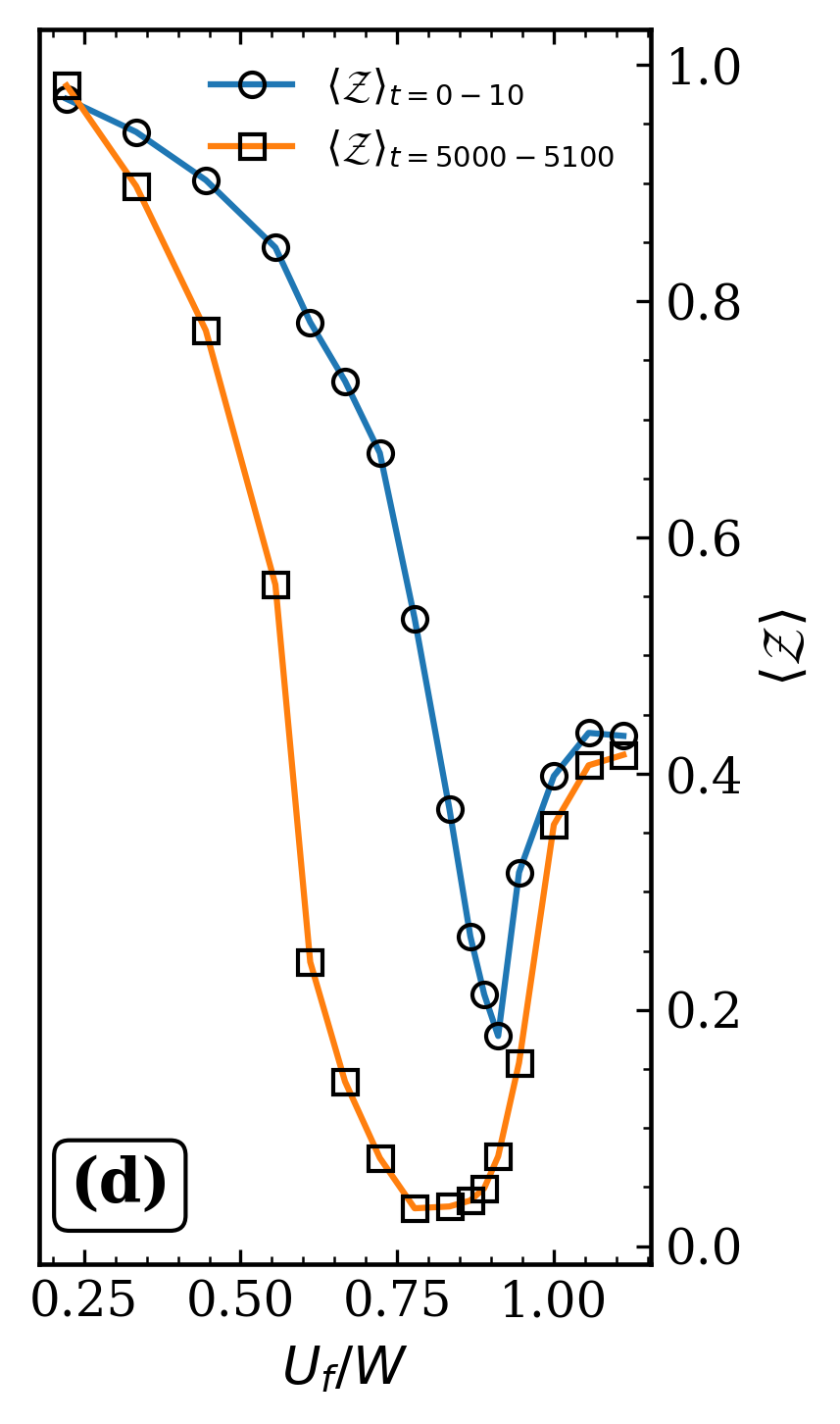}
\caption{%
Dynamics of the site-averaged spectral weight $\mathcal{Z}(t)$ following an
interaction quench to final strength $U_f$.
Panels (a)--(c) show $\mathcal{Z}(t)$ for weak, intermediate, and strong
quenches, respectively.
Panel (d) shows the early-time and long-time averaged spectral weight
$\langle\mathcal{Z}\rangle$ as a function of $U_f/W$.
}

\label{fig:R}
\end{figure}

Using Eq.~(\ref{eq:hga}), the equation of motion for the local Gutzwiller
amplitudes becomes
\begin{align}
i\frac{d\Phi_i}{dt}
&=
U\hat D\Phi_i
-
\sum_{j\sigma} t_{ij}
\frac{\partial R_{i\sigma}}{\partial \Phi_i^\dagger}
R^\ast_{j\sigma}\rho_{ji}
-
\sum_{\sigma}\mu_{i\sigma}\hat N_\sigma\Phi_i
\nonumber\\
&=
\frac{\partial}{\partial \Phi_i^\dagger} H_{\mathrm{sb}},
\label{eq:phi_final}
\end{align}
which closes the coupled dynamics of quasiparticles and local correlation
fields.

In this work we restrict to paramagnetic states, imposing
\begin{align}
n_{i\uparrow}=n_{i\downarrow} &\equiv n_i, \nonumber\\
\Phi_{i0}=e_i,\qquad
\Phi_{i\uparrow}=\Phi_{i\downarrow} &= p_i,\qquad
\Phi_{i\uparrow\downarrow}=d_i.
\end{align}

leading to
\begin{equation}
R_i
=
\frac{
e_i^\ast p_i
+
p_i^\ast d_i
}
{\sqrt{n_i(1-n_i)}}.
\end{equation}

The TDGA equations possess a local gauge freedom associated with the phase of
$\Phi_i$. For paramagnetic states the chemical potentials satisfy
\begin{equation}
\mu_{i\uparrow}=\mu_{i\downarrow}\equiv \mu_i,
\end{equation}
with the explicit expression \cite{SB3}
\begin{equation}
\mu_i
=
\frac{n_i-\tfrac{1}{2}}{n_i(1-n_i)}
\,\mathrm{Re}\!\left(\Lambda_i R_i\right),
\end{equation}
where
\begin{equation}
\Lambda_i=\sum_j -t_{ij}R_j^\ast\rho_{ij}.
\end{equation}

\section{Numerical implementation}
\label{app:dt}
In numerical simulations, Eqs.~(\ref{eq:rho_dyn}) and
(\ref{eq:phi_final}) are integrated simultaneously in time, ensuring full
self-consistency between quasiparticle motion and local correlation dynamics.
This real-space implementation constitutes the Gutzwiller--von Neumann
dynamics method used throughout the main text.

The Hamiltonian contains two independent energy scales: the nearest-neighbor
hopping $t_{\rm nn}$ and the onsite interaction $U$. We set $t_{\rm nn}=1$,
thereby measuring time in units of $\hbar/t_{\rm nn}$, and normalize the interaction
strength by the uncorrelated bandwidth $W=9t_{\rm nn}$ appropriate for the triangular
lattice. All results are therefore reported in terms of $U/W$ and dimensionless
time $t$, unless mentioned otherwise.

The coupled equations of motion are solved using a fourth-order Runge--Kutta
scheme with timestep $\Delta t$. To verify numerical stability and convergence,
we perform simulations for $\Delta t=0.001$, $0.005$, and $0.01$ on a small system of size $12\times12$. As shown in
Fig.~\ref{fig:dt}, these choices yield quantitatively similar trajectories for
the observables reported in the main text.

To seed spatial dynamics without explicitly breaking symmetries, we introduce a
weak on-site Anderson disorder potential $\{\epsilon_i\}$, whose magnitude is
varied across simulations while keeping $\epsilon/W\sim 10^{-6}$. The results
reported are insensitive to the specific realization or precise amplitude of
this weak disorder.

\begin{figure}[t!]
    \centering
    \includegraphics[width=\linewidth]{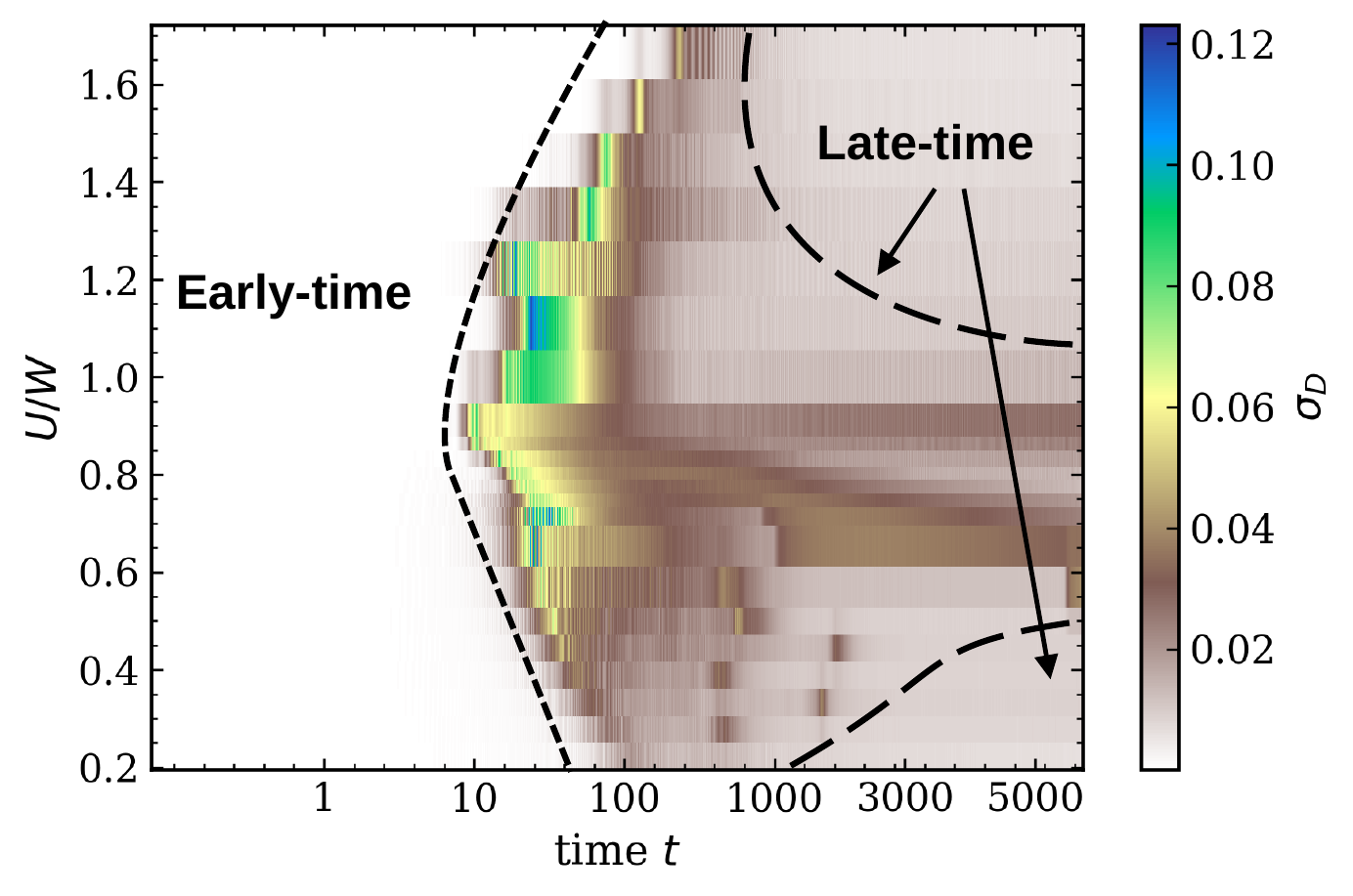}
\caption{%
Color map of the spatial fluctuation of double occupancy,
$\sigma_{\mathcal{D}}(t)$, as a function of time $t$ and interaction strength
$U_f/W$. Dashed lines indicate the approximate boundaries separating the early-
and late-time homogeneous regimes from the intermediate-time inhomogeneous regime.
}
\label{fig:time}
\end{figure}

\section{Spectral Weight}
\label{app:SW}

Within the Gutzwiller approximation, correlation effects renormalize the
hopping amplitudes according to
$
t_{ij} \;\rightarrow\; t_{ij}\, R_i R_j^\ast ,
$
where the site-dependent factors $R_i$ encode the reduction of quasiparticle
coherence due to local correlations. A useful global measure of this
renormalization is the site-averaged spectral weight,
$
\mathcal{Z}(t)=\overline{|R_i(t)|^2},
$
which directly reflects the effective bandwidth of the correlated
quasiparticles.

Figure~\ref{fig:R} shows the time evolution of $\mathcal{Z}(t)$ following
interaction quenches of varying strength. At early times, $\mathcal{Z}(t)$
exhibits coherent oscillations that closely follow the predictions of
homogeneous TDGA, consistent with the behavior of the double occupancy
discussed in the main text. At longer times, however, $\mathcal{Z}(t)$ departs
from the homogeneous TDGA dynamics, reflecting the growth of spatial
inhomogeneities and the breakdown of synchronized quasiparticle motion.

Panel~\ref{fig:R}(d) summarizes the early- and long-time averaged spectral
weight as a function of $U_f/W$. The pronounced deviation between the two
regimes mirrors the behavior observed for the double occupancy, further
confirming that spatial dephasing and inhomogeneity lead to an effective
renormalization of quasiparticle coherence beyond homogeneous descriptions.

\begin{figure}[t]
    \centering
    \includegraphics[width=\linewidth]{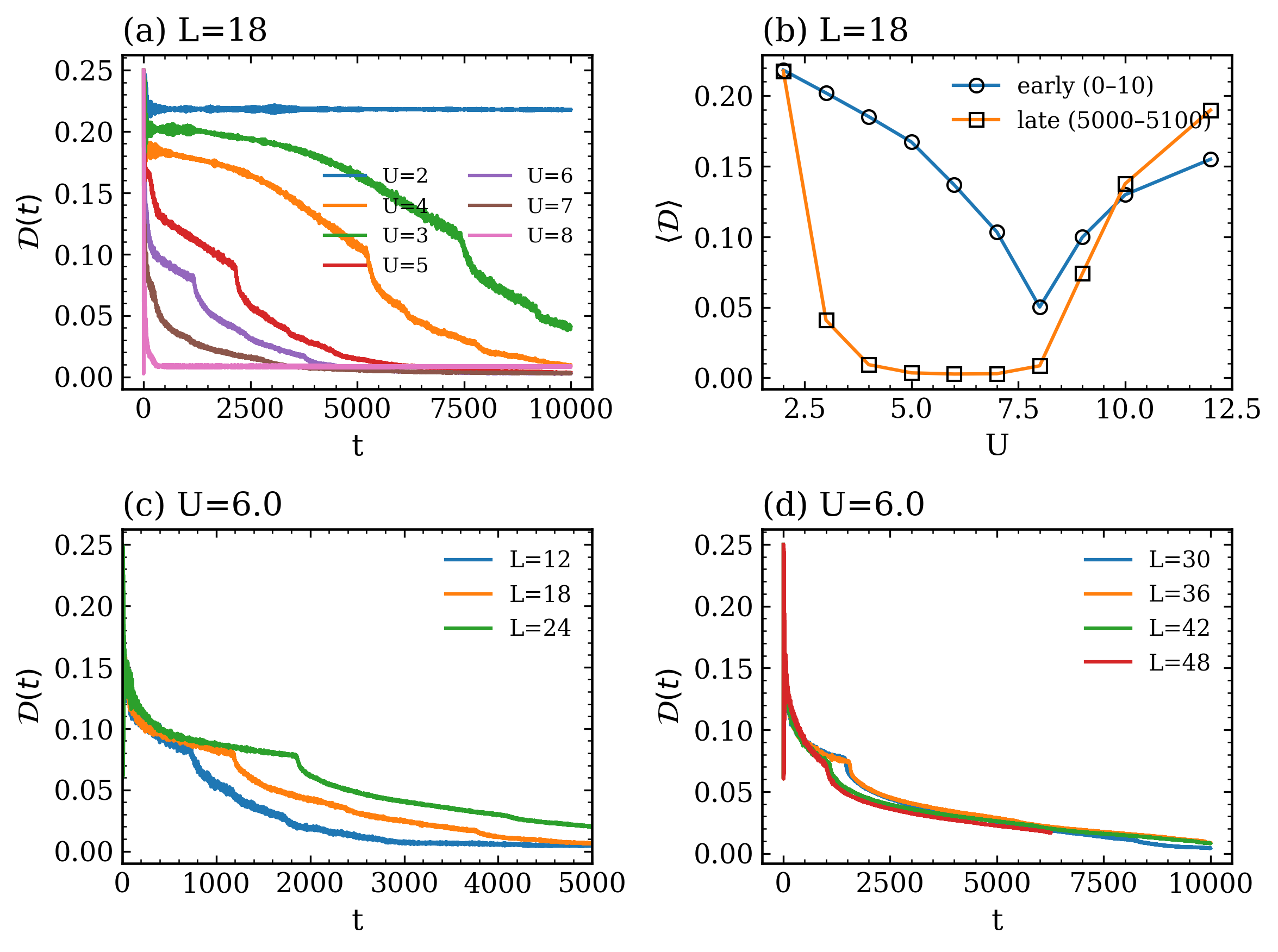}
\caption{%
System-size dependence of the double occupancy dynamics.
(a) $\mathcal{D}(t)$ for an $18\times18$ lattice at different interaction
strengths.
(b) Early-time ($t\in[0,10]$) and late-time ($t\in[5000,5100]$) averages of
$\mathcal{D}$ versus $U_f$.
(c),(d) Size dependence of $\mathcal{D}(t)$ near intermediate coupling,
showing faster relaxation for smaller systems and weak size dependence for
$L\gtrsim30$.
}
    \label{fig:size}
\end{figure}

\section{Temporal regimes}
\label{app:regimes}

Figure~\ref{fig:time} shows the time evolution of the spatial fluctuation of
double occupancy, $\sigma_{\mathcal{D}}(t)$, as a color map for varying
interaction strength $U_f/W$. This representation highlights the separation
between distinct temporal regimes in the post-quench dynamics. At early times,
$\sigma_{\mathcal{D}}(t)$ remains vanishingly small for all interaction strengths,
indicating nearly homogeneous dynamics consistent with the predictions of
spatially uniform TDGA. The extent of this early-time regime is strongly
interaction dependent, as indicated by the dashed boundary.

At longer times, spatial inhomogeneities develop in an interaction-dependent
manner. For weak and strong quenches, $\sigma_{\mathcal{D}}(t)$ approaches a
small but finite value that remains approximately time independent at late
times, as marked by the long dashed guide lines. In contrast, for intermediate
interaction strengths, $\sigma_{\mathcal{D}}(t)$ continues to evolve over the
entire simulation window and does not reach an apparent asymptotic value. This
persistent growth reflects the slow nucleation and coarsening dynamics
discussed in the main text.
\section{System-size effects}
\label{app:size}

Figure~\ref{fig:size} illustrates the dependence of the post-quench dynamics on
system size. Panel~(a) shows the time evolution of the spatially averaged double
occupancy $\mathcal{D}(t)$ for an $18\times18$ lattice over long times
($t\le 10^4$) and for several interaction strengths. While the initial decay and
early-time behavior are similar across different $U_f$, the long-time evolution
exhibits a pronounced dependence on interaction strength.

This behavior is summarized in Fig.~\ref{fig:size}(b), which compares the
early-time and late-time averages of $\mathcal{D}$ as a function of $U_f$. The late-time values
show a markedly different dependence on $U_f$, particularly in the
intermediate-coupling regime where $\mathcal{D}$ remains strongly suppressed.

Panels~(c) and (d) show the size dependence of $\mathcal{D}(t)$ near
intermediate coupling for different linear system sizes $L$. For smaller
systems, the decay of $\mathcal{D}(t)$ occurs more rapidly, while increasing
the system size leads to a slower evolution at long times. For $L\gtrsim30$,
the time dependence of $\mathcal{D}(t)$ becomes comparatively weak within the
accessible simulation time window, indicating reduced finite-size effects in
this regime.

\bibliography{ref}

\end{document}